\documentclass[12pt]{iopart}

\PassOptionsToPackage{prologue,dvipsnames}{xcolor}
\usepackage{bbold}
\usepackage{graphicx} 
\usepackage{bm}
\usepackage{listings}
\usepackage{tikz-cd}
\usepackage{url}
\usepackage[dvipsnames]{xcolor}
\usepackage{enumitem}
\usepackage[sectionbib]{chapterbib}

\usepackage[normalem]{ulem}

\usepackage[caption=false]{subfig}

\usepackage{tabularx}

\definecolor{codegreen}{rgb}{0,0.6,0}
\definecolor{codegray}{rgb}{0.5,0.5,0.5}
\definecolor{codepurple}{rgb}{0.58,0,0.82}
\definecolor{backcolour}{rgb}{0.95,0.95,0.92}

\lstdefinestyle{mystyle}{
  backgroundcolor=\color{backcolour},   commentstyle=\color{codegreen},
  keywordstyle=\color{magenta},
  numberstyle=\tiny\color{codegray},
  stringstyle=\color{codepurple},
  basicstyle=\ttfamily\footnotesize,
  breakatwhitespace=false,         
  breaklines=true,                 
  captionpos=b,                    
  keepspaces=true,                 
  showspaces=false,                
  showstringspaces=false,
  showtabs=false,                  
  tabsize=2,
}

\lstdefinestyle{mystyle1}{
  basicstyle=\scriptsize,
  frame = single,
    showlines=true,
}

\definecolor{codegreen}{rgb}{0,0.6,0}
\definecolor{codegray}{rgb}{0.5,0.5,0.5}
\definecolor{codepurple}{rgb}{0.58,0,0.82}
\definecolor{backcolour}{rgb}{0.95,0.95,0.92}

\lstdefinestyle{mystyle}{
  backgroundcolor=\color{backcolour},   commentstyle=\color{codegreen},
  keywordstyle=\color{magenta},
  numberstyle=\tiny\color{codegray},
  stringstyle=\color{codepurple},
  basicstyle=\ttfamily\footnotesize,
  breakatwhitespace=false,         
  breaklines=true,                 
  captionpos=b,                    
  keepspaces=true,                 
  numbers=left,                    
  numbersep=5pt,                  
  showspaces=false,                
  showstringspaces=false,
  showtabs=false,                  
  tabsize=2,
  showlines=true,
}

\newcommand{\bra}[1]{\langle #1|}
\newcommand{\ket}[1]{|#1\rangle}

\usepackage[linesnumbered,lined,boxed,commentsnumbered]{algorithm2e}

\RestyleAlgo{ruled}
\SetKwComment{Comment}{/* }{ */}

\newtheorem{definition}{Definition}[section]

\newcommand{\argmax}{\mathop{\mathrm{arg\,max}}\limits}

\begin{document}

\title{Context-Aware Unit Testing for Quantum Subroutines}

\author{Mykhailo Klymenko$^1$, Thong Hoang$^2$, Samuel A. Wilkinson$^1$, Bahar Goldozian$^1$, Suyu Ma$^1$, Xiwei Xu$^{2,3}$, Qinghua Lu$^{2,3}$, Muhammad Usman$^{1,4}$, Liming Zhu$^{2,3}$}

\address{$^1$Data61, CSIRO, Research Way, Clayton 3168, VIC, Australia}
\address{$^2$Data61, CSIRO, Level 5/13 Garden St, Eveleigh 2015, NSW, Australia}
\address{$^3$University of New South Wales, Sydney 2052, NSW, Australia}
\address{$^4$School of Physics, The University of Melbourne, Parkville 3010, VIC, Australia}

\ead{mike.klymenko@data61.csiro.au}

\begin{abstract}
Software testing is a critical component of the classical software development lifecycle, and this principle is expected to hold true for quantum software as it evolves toward large-scale production and adherence to industry standards. Developing and testing quantum software presents unique challenges due to the non-deterministic nature of quantum information, the high dimensionality of the underlying Hilbert space, complex hardware noise, and the inherent non-local properties of quantum systems. In this work, we model quantum subroutines as parametrized quantum channels and explore the feasibility of creating practical unit tests using probabilistic assertions, combined with either quantum tomography or statistical tests. To address the computational complexity associated with unit testing in quantum systems, we propose incorporating context-awareness into the testing process. The trade-offs between accuracy, state space coverage, and efficiency associated with the proposed theoretical framework for quantum unit testing have been demonstrated through its application to a simple three-qubit quantum subroutine that prepares a Greenberger–Horne–Zeilinger state, as well as to subroutines within a program implementing Shor’s algorithm.
\end{abstract}

\maketitle

\section{Introduction}

The rapid development of quantum hardware enables the execution of quantum programs of increasing size and complexity (see, for example, IBM’s technology progress report in \cite{ibm_tech}), and brings quantum software into sharp focus.  As a result, monolithic quantum circuits now often include a large number of qubits and operations that can be logically grouped into stand-alone subroutines \cite{10.1145/3656339, klymenko2024architectural}. In the OpenQASM3.0 quantum assembly language \cite{qasm3}, subroutines have been introduced as part of the language syntax \cite{opencasm_subroutines}, not to mention the ability to define subroutines is provided by high-level quantum computing languages such as Cirq, Q\#, and Qiskit. Each subroutine performs a specific task, such as initial state preparations, quantum linear algebra operations, quantum Fourier transform,  the application of parameterized unitary operations, and the introduction of interactions between qubits via entanglement \cite{Nielsen_Chuang_2010, Huang2019, klymenko2024architectural}. 

The emergence of such modular structures and the increasing complexity of quantum circuits make it more difficult to test and debug the entire system. This motivates the development of unit tests for quantum software, much like how the rise of the procedural programming paradigm led to the introduction of unit testing in classical software development \cite{10.1145/800195.805951}. In classical software engineering, unit testing has long been shown as a foundational approach to verify the correctness of individual code modules, enabling early detection of defects, simplified debugging, and improved code maintainability in developing software applications~\cite{zhu1997software, daka2014survey, hamill2004unit}. These principles of unit testing are equally relevant in the quantum domain, as they can help identify faults within complex quantum workflows and mitigate the propagation of errors across the entire quantum circuit.

The empirical unit testing considered in this work aims to detect and isolate both hardware-related failures and human-induced errors in large-scale quantum software systems. To date, the vast majority of research has focused on the validation and verification of quantum hardware rather than software, such as testing of two-qubit gates in silicon quantum dots by evaluating their fidelity using quantum process tomography \cite{Tanttu2024} or assessing of average fidelity for Clifford gates on a 7-qubit quantum processor \cite{PhysRevLett.114.140505}. However, as quantum software systems scale up and quantum hardware becomes increasingly reliable, we anticipate a shift in focus toward addressing human-related errors, as these are expected to persist regardless of advancements in technology. Today, a growing body of literature in quantum software engineering highlights that testing is - or should be - a critical phase in the quantum software development lifecycle \cite{Weder2020, Weder2022, Dwivedi2024}. However, to the best of our knowledge, no systematic quantitative approach to empirical unit testing of quantum software has been proposed to date, and this work aims to fill this gap.

Quantum computing introduces novel approaches to information, giving rise to a new field of knowledge known as quantum information theory \cite{Wilde_2013, watrous2018theory}. The key distinction from classical information theory is that information is encoded in quantum states, with processing achieved through manipulating these states. Unsurprisingly, this shift necessitates the development of new strategies in quantum software engineering, particularly in the testing and debugging of quantum software. There are two main approaches to unit testing quantum software. The first involves performing measurements on the available output qubits and analyzing the resulting statistical data, without requiring additional quantum resources. The second approach takes extra quantum resources, such as additional quantum states, ancillary qubits, and entanglement, to compute evaluation metrics of the tests. An example of this approach is the so-called swap-test \cite{10463159, 10.1145/3656339}, which is analogous to the use of additional light beams and interference effects in optics to infer the quantum states of photons. Given that NISQ-era quantum computers are limited in both qubit count and circuit depth, we will focus on the first approach in this work. 

In this work, we propose a unified quantum unit testing framework based on quantum information theory, which integrates various testing methods and supports different types of assertion statements with diverse arguments. One possible argument for the assertion is a quantum state, which is described by a density matrix. Recovering complete information on a quantum state solely from measurements can be achieved via quantum tomography \cite{chuang1997prescription, PhysRevLett.86.4195, 10.1145/3188745.3188802, Huang2020}, but this scales poorly in the number of qubits and requires a large number of measurements. In this work, we propose unit testing protocols that leverage accessible contextual information to achieve high confidence in test outcomes at a significantly lower measurement cost compared to full quantum process tomography.

To develop a unified testing framework that incorporates different testing procedures for quantum software, we need to address the following research questions:
\begin{enumerate}[label=\textbf{RQ\arabic*:}, leftmargin=2.1cm]
    \item How can a quantum subroutine be formally modeled within the framework of quantum information theory?
    \item How can probabilistic assertions involving various types of arguments, such as density matrices, observables, and quantum state properties - be evaluated?
    \item How does incorporating contextual information enhance the effectiveness of quantum unit testing?
    \item What is the effect of hardware noise on test outcomes?
\end{enumerate}

To address these questions, in the Methods section, we introduce a formal model of a quantum subroutine, quantum unit test, and contextual information, and outline an approach for evaluating probabilistic assertions within unit tests. In the next section Results, we present our findings from case studies in which various assertions and testing protocols are applied to a three-qubit quantum generating a Greenberger–Horne–Zeilinger state, both with and without contextual information, as well as to subroutines implementing the quantum phase estimation as part of Shor’s algorithm. These evaluations are conducted under both ideal (noise-free) conditions and simulated hardware noise.

\section{Method}

\subsection{Quantum subroutine as a specification of a parametrized quantum channel}

To fully leverage the power of quantum information theory for unit testing, we define quantum subroutines and quantum unit tests using fundamental notions of this theory, such as quantum channels, quantum state space, etc. A source code of a quantum program written in a low-level programming language such as OpenQASM \cite{qasm3} is a list of instructions designed to manipulate a multi-qubit quantum state of a quantum register. These instructions must be mapped onto an instruction set specified by the instruction set architecture \cite{smith2016practical} of a specific quantum computer. Each of these operations, including single-qubit unitary gates, multi-qubit gates, and measurements, as well as their compositions, can be mathematically represented by a linear, completely positive trace-preserving map from a space of density matrices into itself, i.e. a quantum channel (also known as the quantum dynamical map or quantum operation) \cite{Wilde_2013}. In the general case, this map can depend on a vector of parameters, $\bm{\theta}$, which can be viewed as classical arguments for quantum subroutines, with the subroutines formally defined as follows:

\begin{definition}[Quantum subroutine]
A quantum subroutine $\Pi$ is a specification given by a source code of a quantum channel $\Phi_{\bm{\theta}}$ that depends on a vector of parameters $\bm{\theta}$:
\begin{equation}
X \stackrel{\Phi_{\bm{\theta}}}{\longrightarrow} Y.
\end{equation}
where $X$ is the Hilbert space of the density matrices for the input quantum register of size $n$, and $Y$ is the Hilbert space of the density matrices for the output quantum register of size $m$. 
\label{def:prog}
\end{definition}

\begin{figure*}[t]
\centering
\subfloat[]{
\includegraphics[height=0.157\linewidth]{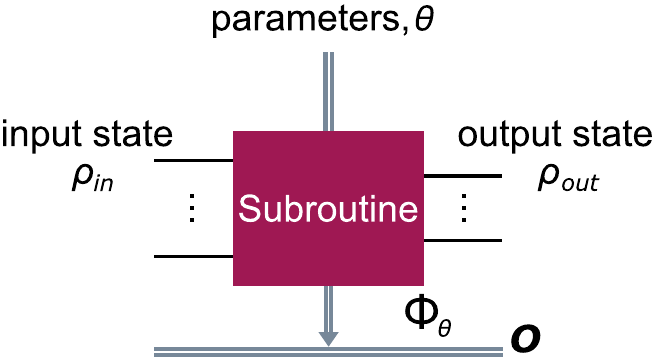}
}\hspace{0.1cm}
\subfloat[]{
\includegraphics[height=0.157\linewidth]{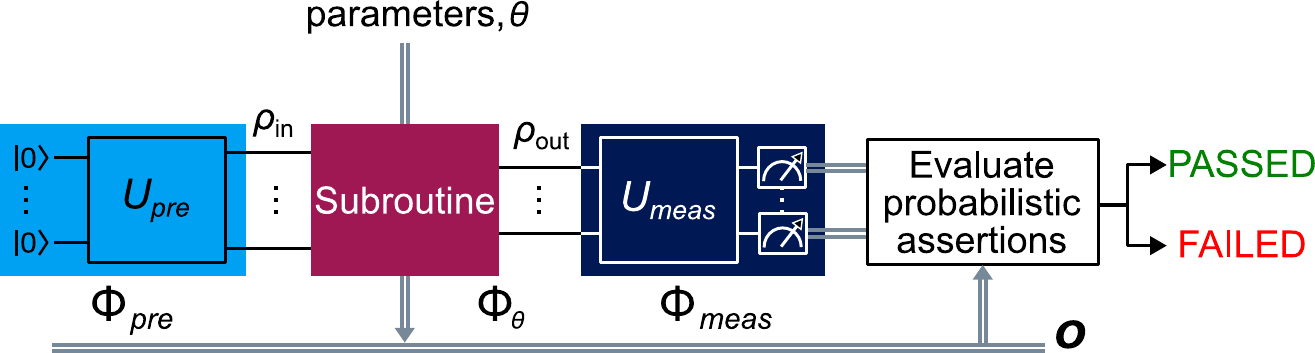}
}
    \caption{(a) Circuit diagram of a general quantum subroutine with both classical and quantum inputs, $\bm{\theta}$ and $\rho_{in}$, and classical and quantum outputs, $\bm{O}$ and $\rho_{out}$. The classical input and output are vectors of real numbers. (b) Circuit diagram of a quantum unit test program composed of the quantum channel $\Phi_{pre}$, which prepares the initial state $\rho_{in}$; the quantum channel $\Phi_{\bm{\theta}}$, representing formally quantum logic of the specified subroutine; the quantum channel $\Phi_{meas}$, which rotates the output state and performs projective measurements; and a classical oracle that analyzes the measurement results and evaluates the assertion statement. }
    \label{fig:sub}
\end{figure*}

In the following, we denote an arbitrary element of the set $X$ by $\rho_{in} \in X$, and an element of the set $Y$ by $\rho_{out} \in Y$. The dimensionality of the spaces $X$ and $Y$ are $4^n$ and $4^m$ correspondingly. A circuit diagram of a quantum subroutine is shown in Figure \ref{fig:sub}a. In the general case, the sizes of the input and output registers may not be equal, $n \neq m$.  When the use of ancillary qubits takes place, the dimensionality of the space $X$ can be smaller than the dimensionality of $Y$.  Also, the dimensionality of the space $Y$ can be smaller than the dimensionality of $X$, if we read out data only from certain qubits, ignoring the states of the others. It is important to note, that the parameters $\bm{\theta}$ can be a vector of real numbers that represent rotational angles for unitary operations, or a classical output of another quantum channel. The latter case defines so-called classical-feedforward \cite{feedforward, Sakaguchi2023}, which refers to the case when one measures states of some qubits and then perform additional quantum operations that depend on the measurement outcome.

A practical specification of a quantum subroutine, defined as a quantum channel, can be realised through various implementations, such as circuit diagrams, quantum programming languages, or matrix representations.

We can distinguish a quantum subroutine from a quantum program based on the following criterion: unlike a quantum subroutine, a quantum program has fixed, well-defined values for all its quantum registers and parameters. Therefore, the definition of a quantum subroutine can be used to define a quantum program as a special case, by specifying the values of its input arguments. Specifically, a quantum program $\Pi$ is a specification of the quantum channel $\Phi_{\bm{\theta}=const}(\rho_0)$, where $\rho_0$ denotes a quantum state of the input quantum register. A quantum program can be represented as a composition of multiple quantum subroutines operating sequentially or in parallel. For instance, it may be expressed as $\Phi_{\bm{\theta}}^1 \circ \Phi_{\bm{\theta}}^2 \circ \left( I(\rho_a) \otimes \Phi_{\bm{\theta}}^3(\rho_0) \right)$, where $I$ denotes the identity channel, $\rho_a$ is the quantum register of ancillary qubits, each $\Phi_{\bm{\theta}}^j$ represents the quantum channel corresponding to the $j$-th quantum subroutine, and the operation $\circ$ is the composition of two linear maps such that $(A \circ B) (x) = A(B(x))$.
 
Defining a program as a quantum channel is consistent with the definition of the Quantum Abstract Machine \cite{smith2016practical} and covers existing low-level instruction set architectures, which include single- and multiple-qubit unitary gate operations, including parametrized ones, measurements, reset operations, and unidirectional data flow \cite{qasm}. Moreover, in \cite{LONG2024112000}, authors report on black-box testing and use a similar definition for quantum programs, pointing out that quantum channels can also encompass possible \textit{if-statements} and \textit{while-loop-statements}. Additionally, the authors of \cite{10.1145/2049706.2049708} have proposed the quantum \textit{while}-language, which is proven to be universal. The semantic function of a quantum \textit{while}-program is a mapping from the program's input state to its output state after execution \cite{10.1145/3428218}. This is consistent with our definition of a quantum program as a parametrized quantum channel.

For further analysis, it is important to distinguish the source code of a quantum program (description of the noise-less linear completely positive trace-preserving map) from the program executed on real devices, which are characterized by noise. To address this, the notion of a quantum computing process modelled as a quantum channel with noise is introduced here:

\begin{definition}[Quantum computing process]
A quantum computing process $\Pi'$ is the instance of the quantum program $\Pi$ that has been deployed and is being executed on a quantum computer characterised by some decoherence rate and noise introducing modifications into the specified quantum channel, $\Phi_{\bm{\theta}} \mapsto \Phi_{\bm{\theta}}'$: 
\begin{equation}
X \stackrel{\Phi_{\bm{\theta}}'}{\longrightarrow} Y.
\end{equation}
\label{def:proc}
\end{definition}

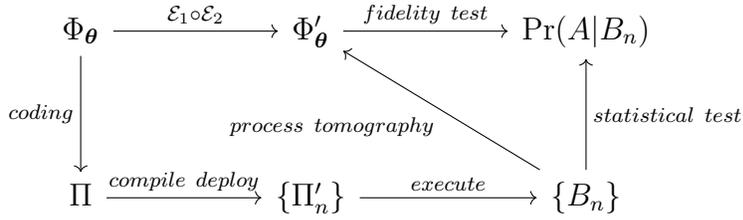
\begin{figure*}
\centering
\begin{tikzcd}[row sep=huge, column sep = huge]
\Phi_{\bm{\theta}} \arrow[d, "coding"'] \arrow[r, "\mathcal{E}_1 \circ \mathcal{E}_2"] & \Phi_{\bm{\theta}}' \arrow[r, "fidelity\ test"] & \mbox{Pr}(A \vert B_n) \\
\Pi \arrow[r, "compile\ deploy"]                                                           & \{ \Pi_n' \} \arrow[r, "execute"] & \{B_n \}\arrow[lu, "process\ tomography"] \arrow[u, "statistical\ test"']
\end{tikzcd}
\caption{\label{diag} Relationships between quantum program lifecycle phases and concepts of quantum information theory.}
\end{figure*}

Distinguishing $\Pi$ from $\Pi'$ indicates that the actual behaviour of the executed code may differ from the specification of the quantum program due to hardware noise and errors. For instance, when the quantum program outputs a pure state according to its specification, the quantum process may still output a slightly mixed state due to the interaction with the environment.

These definitions establish connections between the phases of the quantum program lifecycle and concepts from quantum information theory, as illustrated in the diagram in Fig. \ref{diag}. The abstract representation of the software's functionality is based on the concept of a quantum channel $\Phi_{\bm{\theta}}$. Using programming languages or quantum circuit diagrams, programmers write down a set of instructions (source code of the program $\Pi$) that formally represent the desired functionality. The program $\Pi$ is then compiled and deployed onto a quantum computing device, which initiates one or many instances of the quantum computing process $\Pi'$. One source code can trigger the launch of multiple quantum computing processes, either identical or modified, executed sequentially or in parallel, depending on the parameters set by the compiler, debugger, middleware, and/or operating system. The successful termination of these processes (everything compiles and we do not get any disruptive errors) produces output results $B_n$, where $n$ represents the number of the copies of $\Pi'$. In principle, the compiler can be designed to generate a variety of quantum computing processes, sufficient to produce the output data needed to recover the underlying quantum channel fully or partially. The most informationally complete version of this procedure is called quantum process tomography and returns the fully-recovered quantum channel $\Phi_{\bm{\theta}}'$. Due to the potential for errors, this quantum channel may differ from the desired quantum channel  $\Phi_{\bm{\theta}}$. Mathematically, this unintentional modifications can be described as the action of a composition of two additional quantum channels $\mathcal{E}_1$ and $\mathcal{E}_2$. The first describes the errors introduced by a programmer, while the second represents hardware noise and hardware errors. If the programmer makes no mistakes, $\mathcal{E}_1$ is the identity channel. Similarly, if the hardware is fully fault-tolerant and error-free, then $\mathcal{E}_2$ is the identity channel. Assuming the hardware is functioning correctly and capable of producing reliable output, the quantum software can be tested by evaluating the probability $P(A|B_n)$ - the probability that an assertion $A$
holds true, given the output data $B_n$ obtained from the quantum computer. We refer to this quantity as the probabilistic assertion. Its evaluation allows us to estimate the extent to which the composition of the channels $\mathcal{E}_1 \circ \mathcal{E}_2$ deviates from the identity channel, thereby detecting whether errors have been introduced by the programmer or hardware. Depending on a particular assertion statement $A$ and the available data $B_n$, the probabilistic assertion can be evaluated either directly from the data using statistical tests or via quantum process tomography, which recovers full information about the quantum channel $\Phi_{\bm{\theta}}'$.

\subsection{Unit testing}

For empirical unit tests, the assertions are evaluated based on the information generated from measurements performed on a quantum computing device. Therefore, the assertions for unit tests include information produced by quantum computing processes (see Definition \ref{def:proc}), which correspond to a collection of specially designed quantum test programs $\left\{ T_1, T_2, ... , T_m \right\}$. In this context, we define quantum unit tests as follows:

\begin{definition}[Quantum unit test] 
\label{def:test}
Given a quantum subroutine $\Pi$, a quantum unit test empirically evaluates the conditional probability $\mbox{Pr}\left( A \vert B_n \right)$, where $A$ is a proposition, associated with the expected behavior of $\Pi$, and $B_n$ is a collection of data, generated by $n$ copies of $m$ quantum computing processes $\left\{ T_1', T_2', ... , T_m' \right\}$. The $j$-th quantum computing process corresponds to the quantum program $T_j$ given by the specification of the following composition of quantum channels: 

\begin{equation}
T_j=\left( \Phi_{meas}^j \circ  \Phi_{\bm{\theta}^j} \circ \Phi_{pre}^j \right) \left(\rho \right),
\end{equation}
where $\Phi_{pre}^j$ represents an instance of the quantum channel that prepares the input states, $\rho_{in}^j=\Phi_{pre}^j \left( \rho \right)$, $\Phi_{\bm{\theta}^j}$ is the quantum channel that is specified by the tested quantum subroutine $\Pi$, and $\Phi_{meas}^j$ denotes an instance the quantum channel that performs destructive measurements required to evaluate the probability of the assertion proposition $A$.
\label{def:unit_test}
\end{definition}

This definition implies that during the testing procedure the quantum channel corresponding to the tested subroutine is `sandwiched' between two additional operations: one that prepares the input states, and the other that implements the measurement protocol (see Figure \ref{fig:sub}b), which complies with the pattern ``arrange-act-assert" \cite{okken2022python}. The resulting composition of quantum channels is the body of the unit test program. The explicit expressions for the superoperators $\Phi_{pre}^j$ and $\Phi_{meas}^j$ are given by a unitary channel and a destructive measurement channel, respectively:

\begin{equation}
\Phi_{pre}^j(\rho)=U_{pre}^j \Lambda \left( \rho \right) U_{pre}^{j \dagger},
\label{pre}
\end{equation}

\begin{equation}
\Phi_{meas}^j(\rho)=\sum_{\alpha} \mbox{tr} \left(\mathcal{P}_{\alpha} U_{meas}^{j} \rho U_{meas}^{j \dagger} \mathcal{P}_{\alpha}^{\dagger} \right) \ket{\alpha} \bra{\alpha} 
\label{meas}
\end{equation}
where $\Lambda \left( \rho \right) = \mbox{tr} (\rho) \ket{0}\bra{0}  $ is the qubit reset channel, $U_{pre}^j$ and $U_{meas}^j$ are unitary transformations that modify the quantum state before and after the action of the quantum subroutine 
 $\Pi$, and $\mathcal{P}_{\alpha}$ is a projection operator for the measurement outcome $\alpha$ in the computational basis. By definition, equation (\ref{meas}) always outputs a diagonal density matrix, with the diagonal elements representing the probabilities of the outcomes. Using the cyclic property of the trace, equation (\ref{meas}) can be rewritten as:

\begin{equation}
\Phi_{meas}^j(\rho)=\sum_{\alpha} \mbox{tr} \left(\mathcal{V}_{\alpha}^j \rho  \right) \ket{\alpha} \bra{\alpha} 
\label{meas1}
\end{equation}
where $\mathcal{V}_{\alpha}^j=U_{meas}^{j \dagger} \mathcal{P}_{\alpha}^{\dagger} \mathcal{P}_{\alpha} U_{meas}^{j}$.

Authors in \cite{9438603, 9678798} propose a notion of test coverage for quantum programs and distinguish three coverage criteria for black-box testing: input coverage, output coverage and input-output coverage. Using the definition of the quantum subroutine (Definition \ref{def:prog}), we can also distinguish state space coverage and parameter space coverage. The former indicates the fraction of the input and output Hilbert spaces, $X$ and $Y$, that is covered by a unit test for a given set of parameters $\boldsymbol{\theta}^j$. The latter refers to the fraction of the parameter space that is covered by unit tests, for any particular combination of input and output states. The coverage depends on the variations of the input states, prepared by the process $\Phi_{pre}^j$, the variations of the parameters $\bm{\theta}^j$, and the variations of the projective measurements described by the process $\Phi_{meas}^j$.  

A particular choice of operators of the set $\Phi_{pre}^j$ and $\Phi_{meas}^j$ determines elements of the set of quantum test programs $\left\{ T_1, T_2, ... , T_m \right\}$ and constitutes a quantum testing protocol. The testing protocol is fully defined by the specification of the sets 
$\{ \Phi_{pre}^j\}$ and $\{ \Phi_{meas}^j\}$, along with all possible pairwise combinations of their elements. 

In this work, we demonstrate that the choice of a quantum testing protocol depends on the type of assertion arguments, the requirements for code coverage and accuracy, and the available contextual information. Some of the possible protocols are discussed in Section \ref{testing_protocols} below and are listed in Table~\ref{tab1}.

\begin{table*}[t!]
    \centering
        \caption{\label{tab1}Testing protocols (the complexity is shown for $n=N \cdot \alpha$, where $N$ is number of qubits, $\alpha$ is number of shots).}
        \footnotesize
\begin{tabularx}{\linewidth} { 
  | >{\raggedright\arraybackslash}p{\dimexpr.18\linewidth-2\tabcolsep-1.3333\arrayrulewidth} 
  |  >{\raggedright\arraybackslash}p{\dimexpr.06\linewidth-2\tabcolsep-1.3333\arrayrulewidth}| >{\raggedright\arraybackslash}p{\dimexpr.11\linewidth-2\tabcolsep-1.3333\arrayrulewidth} | >{\raggedright\arraybackslash}p{\dimexpr.4\linewidth-2\tabcolsep-1.3333\arrayrulewidth} | >{\raggedright\arraybackslash}p{\dimexpr.25\linewidth-2\tabcolsep-1.3333\arrayrulewidth}|  }
  \hline
Protocol  & Refs. & Complex- ity & Context information to achieve full coverage & Possible arguments for assertions  \\ 
    \hline
   Process tomography &\cite{PhysRevA.77.032322}&O$\left( 2^{4n} \right)$ & $\left(C_X=X, C_Y=\mathcal{Y} \right)$, no context required. & Choi matrix, Pauli transfer matrix, Kraus matrix, etc. \\ 
   
    State tomography & \cite{paris2004quantum}& O$\left( 2^{2n} \right)$ & $\left(C_X=\{\rho_{in}\} \subset X, C_Y=\mathcal{Y} \right)$, the input state is fixed or there are few possible input states. & Density matrices \\ 
    
    Classical shadow tomography &\cite{Huang2020}& O$\left( \frac{\log{M}}{ \varepsilon^2 }\right)$ & $\left(C_X=\{\rho_{in}\} \subset X,  C_Y= \{\mbox{tr}\left(O_j \rho \right) \} \right)$, where $O_j$ is an element from the set of $M$ observables, and $ \varepsilon$ is precision. & Classical shadows, observables  \\ 
    
    Statistical tests (Pearson's $\chi^2$) &\cite{Huang2019}& O$\left(n \right)$ & $\left(C_X=\{\rho_{in}\} \subset X, C_Y= \{\mbox{tr}\left(O \rho \right) \} \right)$, where $O$ is the operator corresponding to the projective measurements in the computational basis set. & Statistical characteristics of a quantum state, e.g. expectation value, degree of entanglement, superposition \\ 
    
 Single shot measurement (classical assertions) & & O$\left(N \right)$ &   $\left(C_X=\{\rho_{in}\} \subset X,\right.$ \newline $\left.C_Y= \{\mbox{tr}\left(O \rho \right) \} \subset \{0, 1\} \right)$, where $O$ is the operator corresponding to the projective measurements in the computational basis set. The output is expected to be deterministic. & Deterministic output \\ 
 \hline
\end{tabularx}
\end{table*}

\subsection{Testing protocols}
\label{testing_protocols}

\subsubsection{Protocol based on the standard quantum process tomography}

In order to achieve the maximal state space coverage, one of the possible choices for the set $\{\Phi_{pre}^j \}$ is a matrix basis set that span the space of input density matrices $X$. One of such bases are the Pauli strings.  Indeed, any density matrix $\rho_{in}$ can be expressed as a linear combination of these operators:

\begin{equation}
\rho_{in} =\sum_{j_1,..,j_N} c_{j_1,..,j_N} P_{j_N} \otimes ... \otimes P_{j_1} = \sum_{\mathbf{j}} c^{\mathbf{j}} P^{\mathbf{j}}
\label{pauli}
\end{equation}
where $P_i \in \{I, X, Y, Z \}$ is one of four possible single-qubit Pauli operators (given by so-called Pauli gates in quantum programming languages), $N$ is number of qubits, and $\mathbf{j}$  is a vector of indices with $N$ elements. Therefore, full state space coverage is achieved by the following choice of  $\{\Phi_{pre}^j \}$:
\begin{equation}
\{ \Phi_{pre}^j\} =\{ P^{\mathbf{j}} \circ \Lambda \vert P^{\mathbf{j}} \in Q^N \}
\label{pauli1}
\end{equation}
where $Q^N$ is Pauli basis for the Hilbert space of $N$ qubits, and $\Lambda$  is the qubit reset channel.

Additionally, a set of operators $\{\mathcal{P}_{\alpha}^j\}$ and corresponding set $\{ \Phi_{meas}^j\}$ are chosen to constitute the informationally complete positive operator-valued measure (POVM). This choice  provides sufficient information $B_n$ to reconstruct fully the output state $\rho_{out}$ \cite{PhysRevA.64.052312}. Note that the output density matrix can also be expanded using the Pauli basis:

\begin{equation}
\rho_{out} = \Phi_{\mathbf{\theta}}(\rho_{in}^j)  = \sum_{\mathbf{k}} c^{\mathbf{k}} P^{\mathbf{k}}
\label{pauli_out}
\end{equation}
The coefficients are expectation values of the Pauli operators: $c^{\mathbf{k}}=\left(1 / 2^N \right) \mbox{tr} \left(P^{\mathbf{k}} \rho \right)$. Thus, to obtain all unknown coefficients, the set 
$\{\Phi_{meas}^j \}$ can be chosen as:  
\begin{equation}
\{ \Phi_{meas}^j \} =\{ P^{\mathbf{k}} \vert P^{\mathbf{k}} \in Q^M \}
\label{pauli2}
\end{equation}
where $Q^M$ is Pauli basis for the Hilbert space of $M$ output qubits.

The testing protocol determines the set of the quantum testing programs $\left\{ T_1, T_2, ... , T_m \right\}$ (see Definition \ref{def:unit_test}). Having two sets $\{ \Phi_{meas}^j\}$ and $\{ \Phi_{pre}^j(\rho_0) \}$ defined, these testing programs can be created by forming all possible pairs of elements, with one element taken from each of the two sets. This testing protocol is identical to standard quantum process tomography and recovers full information about the quantum channel \cite{PhysRevA.77.032322}. To prove this, we compute the Frobenius inner product for matrices $\rho_{in}^j$, given by equation (\ref{pauli}), and $\rho_{out}$, given by equation (\ref{pauli_out}). By repeating these computations for all possible pairs from $\{ \Phi_{meas}^j\}$ and $\{ \Phi_{pre}^j \}$, and combining the results into a matrix, we recover the Pauli transfer matrix \cite{hantzko2024pauli}, $PTM$, whose elements are given by: $PTM_{\mathbf{j}\mathbf{k}} = \langle  P^{\mathbf{j}} , c^{\mathbf{k}} P^{\mathbf{k}} \rangle_{F}$. The Pauli transfer matrix fully defines the quantum channel $\Phi_{\mathbf{\theta}}$, and recovering it from the measurement data is, by definition, the standard quantum process tomography. There are other equivalent representations of quantum channels, such as Choi matrices and Kraus operators \cite{watrous2018theory, Wilde_2013}, which can be converted from one to another \cite{hantzko2024pauli}. These matrices are natural arguments for assertion statements of the unit tests. 

While full coverage is an advantage of this protocol, it also has drawbacks that are well-documented in studies of quantum process tomography. First, its computational complexity scales as  O$\left( 2^{4n} \right)$ with the number of qubits. Additionally, complete knowledge of the Pauli transfer matrix may not always be available or, in the case of a large number of qubits, may not even be feasible to store.

\subsubsection{Protocol based on the quantum state tomography}

Another possible testing protocol is based on the quantum state tomography. It is implemented similarly to the one described above except the set of input density matrices contains only one element:

\begin{equation}
\{ \Phi_{pre}^j(\rho) =\rho_{in} \}, \;\rho_{in} \in X,
\end{equation}
This protocol offers limited code coverage, as it provides a test only for a single fixed input quantum state of the subroutine. The set $\{ \Phi_{meas}^j \}$, however, still consists of POVM operators to ensure a tomographically complete set of measurements. The protocol is less computationally demanding and reconstructs the full density matrix for a given input. In this case, the density matrix can be used as an argument for the assertion statements.

\subsubsection{Protocol based on the classical shadow tomography}

Additionally to restricting the space of input states for the testing protocols, one can also reduce the set $\{ \Phi_{meas}^j \}$. One of the systematic ways to do it has been offered by the quantum shadow tomography \cite{10.1145/3188745.3188802, Huang2020}, where the measurements are randomised and performed in an optimal way to recover a set of expectation values of observables $O_j$ related to the density matrix via the relationship $\mbox{tr}\left( O_j \rho \right)$, which can in turn be used for the assertions. Although this protocol cannot recover the whole density matrix, it requires fewer measurements and is therefore significantly less computationally demanding than the two protocols discussed above. 

\subsubsection{Protocol based on statistical tests}

In the extreme case, the set $\{ \Phi_{meas}^j \}$  can consist of a single projective measurement operator, while the set $\{ \Phi_{pre}^j \}$ also has one element -- a channel that prepares the system in a specific quantum state with density matrix $\rho_{in}$. Sampling from a statistical distribution corresponding to this operator can be useful for unit testing, as it may contain information about expectation values or other statistical properties of quantum states. Indeed, the authors in \cite{Huang2019} proposed using the $\chi^2$-test to verify the superposition and entanglement of quantum states. 

The $\chi^2$-test is a statistical hypothesis test commonly used in the analysis of contingency tables, particularly when sample sizes are large. In general, the $\chi^2$-test can serve as the foundation of a testing protocol for quantum software when the assertion statement $A$ (see Definition \ref{def:unit_test}) represents a statistical model, and the proposition that this model adequately describes the outcome frequencies from the quantum computer is treated as the null hypothesis.

\subsection{Evaluation of probabilistic assertions}

\subsubsection{Probabilistic assertions for density matrices and quantum channels}

The practical evaluation of the probabilistic assertion $\mbox{Pr}(A \mid B_n)$ in Definition~\ref{def:test} plays a critical role in unit testing and depends on the specific testing protocol and the types of arguments used -- such as density matrices, Choi matrices, or functionals of density matrices that yield either real-valued outputs (e.g., expectation values of observables or binary outcomes representing specific quantum state properties).

In general, the set $B_n$ denotes a collection of data produced by one or more quantum computing processes, intended to yield measurement counts from either a single projective measurement or multiple projective measurements, depending on the protocol and the specific assertion arguments.

 In this section, we consider predicates $A$ where one of the propositional variables is a density matrix, i.e. $A[\rho]$ is a function of density matrix. We later extend this class of assertions to include cases where the arguments are quantum channels, represented by their corresponding Choi matrices. We apply the law of total probability for conditional probabilities \cite{gelman2013bayesian} to compute the probability that $A$ is true, given $B_n$:

 \begin{equation}
     \mbox{Pr}(A \left\vert\right. B_n) = \int d \rho \; \mbox{Pr}(A[\rho] \left\vert\right. \rho, B_n) \mbox{Pr}(\rho \left\vert\right. B_n)
     \label{bayes}
 \end{equation}
 where $d \rho $ is the measure in the Hilbert space, such that $\int d \rho =1$. The quantity $\mbox{Pr}(\rho \left\vert\right. B_n)$ is well known from the literature on Bayesian quantum state estimation \cite{Granade_2016, Lukens_2020}, and is usually computed using Bayes’ theorem:

\begin{equation}
    \mbox{Pr}(\rho \left\vert\right. B_n) = \frac{~\mbox{Pr}(B_n \left\vert\right. \rho) \mbox{Pr}(\rho)  }{\int d \rho ~ \mbox{Pr}(B_n \left\vert\right. \rho)}
    \label{bayes0}
\end{equation}
The conditional probability function reads $\mbox{Pr}(B_n \left\vert\right. \rho) = \prod_k \left( \mbox{tr} [F_k \rho] \right)^{n}$, where $F_k$ is one of the measurement operators and $n$ is the number of system copies. The density matrix $\rho$ can be reconstructed with an arbitrary precision provided that the set $\{F_k\}$ constitutes a tomographically complete set of POVM operators and an infinite number of measurements are performed for each operator \cite{PhysRevA.64.052312, PhysRevLett.86.4195}. Since, in practice, only a finite number of measurements can be made, reliable quantum state tomography must be characterized in terms of obtained precision, error bars, and other relevant metrics, the importance of which has been recognized early by several research groups \cite{Smolin_2012, Renner_2012, Renato_2016, Renner_2019}. Computing $\mbox{Pr}(B_n \left\vert\right. \rho)$ efficiently over the entire Hilbert space is not an easy task, and several methods have been proposed \cite{Smolin_2012, Renato_2016, Kleinmann_2024} and are still evolving. In what follows, we use the notation $$\mu_{B_n} \left( \rho \right)=\frac{\mbox{Pr}(B_n \left\vert\right. \rho)}{\int d \rho ~ \mbox{Pr}(B_n \left\vert\right. \rho)},$$
which allows to rewrite equation~(\ref{bayes}) as follows:

\begin{equation}
    \mbox{Pr}(A \left\vert\right. B_n)  = \int d \rho ~  \mbox{Pr}(A[\rho] \left\vert\right. \rho, B_n) \; \mu_{B_n} \left( \rho \right)  \mbox{Pr} \left( \rho
    \right).
    \label{posterior}
\end{equation}
Note, that $\int d \rho~\mu_{B_n} \left( \rho \right)  =1$. The quantity $\mu_{B_n}$ was introduced in \cite{Renner_2012} in the context of estimating the quality of the quantum state tomography and is called the estimate density. The function $\mu_{B_n} \left( \rho \right)$ has maximum located at $\rho_{B_n}$:

$$\rho_{B_n} = \argmax_{\rho \in Y}~\mu_{B_n}(\rho ).$$

Computing the estimate density is a non-trivial task, but some cases it can be found analytically. Let us consider two such cases:

\begin{enumerate}
    \item The set of measurements is tomographically complete and an extremely large number of samples is taken making the distribution $\mu_{B_n}(\rho)$ extremely narrow: $$\mu_{B_n} \left( \rho \right)  \sim \delta\left(\rho - \rho_{B_n} \right),$$ where $\delta(\cdot)$ is a delta function.
    \item The distribution $\mu_{B_n}(\rho)$ varies with $\rho$ much faster than the function $\mbox{Pr}(A[\rho])$. In this case, the function $\mbox{Pr}\left( A[\rho] \right)$ can be approximated by a constant $\mbox{Pr}\left( A[\rho_{B_n}] \right)$.
\end{enumerate}

Both these cases significantly simplify the assertion probability in equation~(\ref{posterior}) resulting in:
\begin{equation}
    \mbox{Pr}(A \left\vert\right. B_n)  \approx  \mbox{Pr}\left( A[\rho=\rho_{B_n}] \left\vert\right. \rho_{B_n}, B_n  \right) \mbox{Pr} \left( \rho_{B_n}
    \right).
    \label{posterior0}
\end{equation}

Since $\mu_{B_n}(\rho)$ is typically described by Gaussian models \cite{PhysRevA.64.052312, Smolin_2012, Renner_2012, Renato_2016, Kleinmann_2024}, which are characterized by exponentially decaying tails, the second case is quite realistic, assuming that quantum state tomography has been performed reliably. The author of \cite{JONES1991140} proposes using a normalized uniform integration measure as the prior -- by setting $\mbox{Pr} \left( \rho \right)=1$ -- in cases where no prior information about the distribution of quantum states is available. This assumption further simplifies equation~(\ref{posterior0}):

\begin{equation}
    \mbox{Pr}(A \left\vert\right. B_n)  \approx  \mbox{Pr}\left( A[\rho=\rho_{B_n}] \left\vert\right. \rho_{B_n}, B_n  \right).
    \label{posterior1}
\end{equation}

From here on in this work, we therefore assume that equation~(\ref{posterior1}) holds as an approximation, however, for more precise estimates, the estimate density $\mu_{B_n}(\rho)$ should be explicitly computed.

Let us consider a particular example of the predicate $A$: $\rho = \rho_{expected}$, which expresses equivalence of the output density $\rho$ matrix to some expected density matrix $\rho_{expected}$. Since the predicate $A$ does not explicitly depend on $B_n$, equation~(\ref{posterior1}) simplifies to:

\begin{equation}
    \mbox{Pr}(A \left\vert\right. B_n)  = \mbox{Pr}\left( \rho_{expected}=\rho_{B_n} \left\vert\right. \rho_{B_n}\right).
    \label{posterior2}
\end{equation}

This conditional probability measures the distinguishability between the expected state $\rho_{expected}$ and experimentally realized state $\rho_{B_n}$. The equivalence of density matrices (or quantum states), considering probabilistic nature of the information they represent, implies statistical equivalence. The statistical equivalence means that, given data samples generated from one of two defined density matrices, we cannot determine which density matrix was used to generate data. This relates to the problem of discrimination of quantum states \cite{fuchs1996distinguishability, Audenaert, Bae_2015} and the problem of quantum hypothesis testing \cite{Kargin2005, Nussbaum2009, hayashi2006quantum, tomamichel2015quantum}, which are also connected to the distinguishability of classical probability distributions \cite{Chernoff}. In this context, the probability in equation~(\ref{posterior2}) reads $\mbox{Pr}\left( \rho_{expected}=\rho \left\vert\right. \rho \right) = 2P_{e,min}$, where $P_{e,min}$ represents the minimum probability of error in distinguishing between two quantum states, which according to \cite{Audenaert, tomamichel2015quantum} reads:

\begin{equation}
 P_{e,min} = \frac{1}{2} \left(1 - \big\| \rho-\rho_{expected} \big\|_{tr} \right),
 \label{trace}
\end{equation}
where $\big\| \cdot \big\|_{tr}$ is the trace norm.

This expression implies that the states with density matrices $\rho$ and $\rho_{expected}$ are indistinguishable when the probability of error is 0.5, indicating maximal uncertainty in discriminating these states. In this case, the probabilistic assertion returns one: $\mbox{Pr}(\rho =\rho_{expected}) = 1$. 

Using the trace norm as a measure of distinguishability of quantum states in equation~(\ref{trace}) has issues related to its non-monotonic dependence on the number of copies of the system \cite{Audenaert}. The equality in equation~(\ref{trace}) can be relaxed by replacing it with an inequality that includes an upper bound from the Fuchs–van de Graaf inequalities \cite{Audenaert, tomamichel2015quantum}:

$$1 - \big\| \rho-\rho_{expected} \big\|_{tr}  \leq F(\rho,\rho_{expected}),$$
where $F(\rho,\rho_{expected})$ is fidelity of quantum states.

In what follows, we approximate the exact value by the upper bound, leading to the final expression for the probabilistic assertion of the equality of two density matrices:

\begin{equation}
    \mbox{Pr}( \rho_{expected} = \rho \left\vert\right. \rho ) \sim F(\rho,\rho_{expected}).
    \label{prior}
\end{equation}

Combining equation~(\ref{posterior2}) and equation~(\ref{prior}), the probabilistic assertion is given by:

\begin{equation}
    \mbox{Pr}( \rho_{expected} = \rho_{B_n} \left\vert\right. \rho_{B_n}) \sim F(\rho_{B_n},\rho_{expected}).
    \label{eq:fid}
\end{equation}

Note that equation~(\ref{eq:fid}) can be used to estimate quantum channel equivalence, not just for density matrices. This is achieved by replacing the density matrices in equation~(\ref{eq:fid}) with the normalized Choi matrices $C$:

\begin{equation}
    \mbox{Pr}( C_{expected} = C_{B_n} \left\vert\right. C_{B_n}) \sim F(C_{B_n}, C_{expected}).
    \label{eq:fid1}
\end{equation}
where $F(C_{B_n}, C_{expected})$ is the quantum process fidelity \cite{PhysRevA.93.042316}. This transition is theoretically justified by the channel-state duality principle, which relates quantum states and channels through the Choi–Jamiołkowski isomorphism \cite{Wilde_2013}.

\subsubsection{Probabilistic assertions for density matrix functionals and properties of quantum states}

Another type of predicate $A$ considered in the assertions examined in this work involves the equivalence of a functional of a density matrix to a specified value $f[\rho] = p_{expected}$. The functional $f[\rho]$ is defined as the map: $X \rightarrow \mathbb{R}$. Examples of such functionals include the operator expectation value, $f[\rho] = \mbox{tr}\left[ \hat{O} \rho \right]$, purity of the quantum state, $f[\rho] = \mbox{tr}\left[ \rho^2 \right]$, etc. 

Functionals of the density matrix are often related to statistical properties that can be directly estimated from the measured data $B_n$, without requiring full knowledge of the density matrix. In this case, equation~(\ref{posterior1}) from the previous section does not explicitly depend on the density matrix, but rather only on the observed data:

\begin{equation}
    \mbox{Pr}(A \left\vert\right. B_n)  =  \mbox{Pr}\left( f[\rho_{B_n}] = p_{expected} \left\vert\right. B_n  \right).
    \label{property_assertion}
\end{equation}

To practically evaluate the probabilistic assertion in equation~(\ref{property_assertion}), we need to employ a metric that evaluates the distinguishability between the expected probability distribution and the probability distribution inferred from the measured frequencies of outcomes related to the functional values $f[\rho_{B_n}]$. Statistics and information theory offer several methods and metrics to evaluate the distinguishability of probability distribution functions, including the Chernoff bound, the $\chi^2$-test \cite{Huang2019} (which have been used for entanglement or purity tests in \cite{Huang2019}), and others. 

In this paper, we follow the methodology proposed in \cite{Huang2019} and interpret equation~(\ref{property_assertion}) as evaluating the goodness of fit between the count frequencies obtained from measurements and the expected distribution, using the 
Pearson's $\chi^2$-test \cite{vaughan2013scientific}. In this context, the resulted probability is given by the p-value inferred from the test \cite{vaughan2013scientific}.

Properties of quantum states, being defined in \cite{montanaro2013survey},  are similar to density matrix functionals, except that instead of returning a real number, the functional returns a binary value, 0 or 1, that indicates whether the state possesses a given property. For more details, see the definition of the property notion in \cite{montanaro2013survey}. In addition to reducing computational complexity, property-based tests are particularly useful when complete information about the expected density matrix, $\rho_{expected}$, is either unknown or impractical to store due to its high dimensionality.

\subsection{Introducing context}

In general, contextual information for quantum unit testing can be divided into three categories:

\begin{enumerate}[label=\arabic*.]
    \item Input scope: Defines the subspace of the input density matrix space and the possible input parameter values if any.
    \item Output usage: Concerns how the output states will be utilized within the software. 
    \item General context: Includes other types of contextual information, such as the probability of the user making certain types of errors.
\end{enumerate}

Each type of contextual information can be either deterministic or probabilistic. In this work, we focus solely on the deterministic contextual information, particularly on the input scope and output usage, hereafter referred to simply as 'context,' which can be formally defined as:

\begin{definition}[Context]
The context is defined as a tuple $C = (C_X, C_Y)$, where:
\begin{itemize}
    \item $C_X \subset X$ is a (possibly countable or uncountable) subset of the Hilbert space $X$, specifying the input domain of the subroutine, and
    \item $C_Y$ is a collection of Hermitian operators or density matrix functionals that describe the intended usage of the subroutine's output state.
\end{itemize}
In the absence of restrictions on the intended usage, this is expressed by setting 
$C_Y = \mathcal{Y}$, where 
$\mathcal{Y}$ denotes the unrestricted set of allowable output usages.
\end{definition}

The context defined in this form describes the restrictions placed on the domain of input and output density matrix processes within a quantum subroutine. Its purpose is to reduce the computational complexity of the unit tests required for achieving full coverage. For example, the tested unit may be part of a larger program that operates on density matrices within a restricted Hilbert space (such as a specific hyperplane) or even within a more limited domain of this hyperplane (e.g., when parameters defining rotations take specific values). The context may also specify that only partial information from the density matrix output by the tested unit will be used in subsequent computations. Let us consider some examples of the usage of this formalism. If we know that the input state is always represented by the density matrix $\rho$, the context takes the form:   $C = (C_X=\{\rho\}, C_Y=\mathcal{Y})$. More examples of contextual information are shown in Table ~\ref{tab1}.

\section{Results}

\subsection{Experimental setup}

\begin{figure}[t]
\centering
\includegraphics[width=0.45\linewidth]{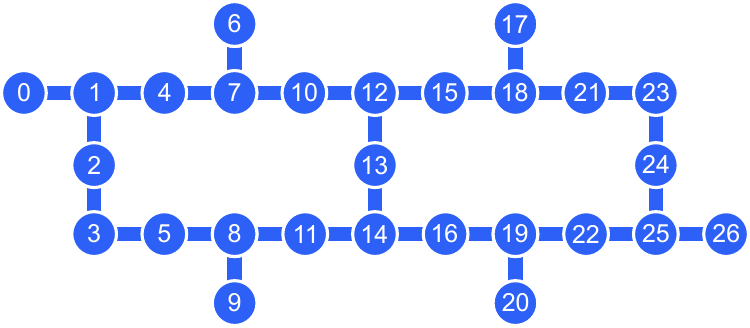}
\caption{\label{topology}Experiments have been conducted on a simulator replicating 27 qubit IBM backend \textit{ibm\_sydney}.}
\end{figure}

\newsavebox{\firstlisting}
\begin{lrbox}{\firstlisting}
\begin{minipage}[t]{0.23\textwidth}
\centering
\begin{lstlisting}[language=Python]
//sample1

def sample1(qubit[3] q){

x q[0];
h q[0];

cx q[0], q[1];
cx q[0], q[2];
}

\end{lstlisting}
 \includegraphics[height=0.5\textwidth]{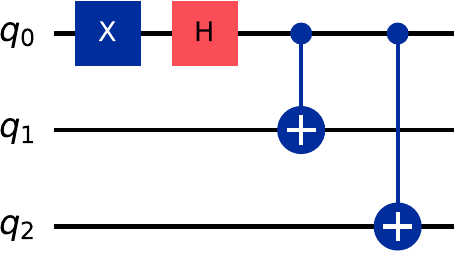}
\end{minipage}
\end{lrbox}

\newsavebox{\secondlisting}
\begin{lrbox}{\secondlisting}
\begin{minipage}[t]{0.23\textwidth}
\centering
\begin{lstlisting}[language=Python]
//sample2

def sample2(qubit[3] q){

x q[0];
h q[(*@\fontfamily{qcr}\scriptsize \textcolor{red}{1}@*)];

cx q[0], q[1];
cx q[0], q[2];
}

\end{lstlisting}
 \includegraphics[height=0.5\textwidth]{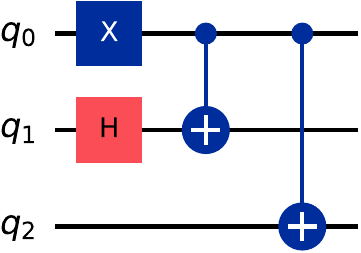}
\end{minipage}
\end{lrbox}

\newsavebox{\thirdlisting}
\begin{lrbox}{\thirdlisting}
\begin{minipage}[t]{0.23\textwidth}
\centering
\begin{lstlisting}[language=Python]
//sample3

def sample3(qubit[3] q){

(*@\fontfamily{qcr}\scriptsize  \textcolor{red}{s}@*) q[0];
h q[0];

cx q[0], q[1];
cx q[0], q[2];
}

\end{lstlisting}
 \includegraphics[height=0.5\textwidth]{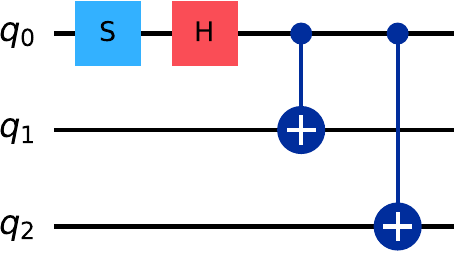}
\end{minipage}
\end{lrbox}

\newsavebox{\forthlisting}
\begin{lrbox}{\forthlisting}
\begin{minipage}[t]{0.23\textwidth}
\centering
\begin{lstlisting}[language=Python]
//sample4

def sample4(qubit[3] q){

x q[0];
h q[0];

(*@\fontfamily{qcr}\scriptsize  \textcolor{red}{cx q[1], q[2];}@*)
cx q[0], q[1];
cx q[0], q[2];
}
\end{lstlisting}
 \includegraphics[height=0.5\textwidth]{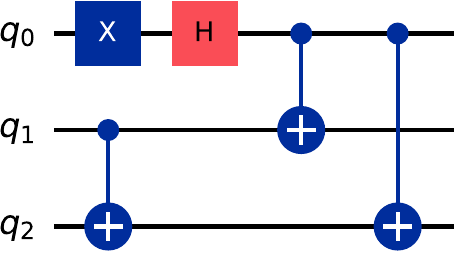}
\end{minipage}
\end{lrbox}

In this section, we conduct a series of numerical experiments to gather empirical data on the performance of different testing protocols and to examine how the test results depend on the number of measurements and the available contextual information. All experiments were carried out using a simulator called \texttt{FakeSydneyV2} \cite{FakeSydneyV2}, which is part of the \texttt{Qiskit} classes library \cite{qiskit} from IBM. This simulator replicates the IBM backend provided by 27-qubit IBM quantum processing unit \texttt{ibm\_sydney} with the connectivity of qubits shown in Figure \ref{topology}. The experiments were performed in both a noise-free environment and an environment with simulated noise. In the latter case, a noise model was generated from system snapshots obtained in the past by observing the behavior of a real system. The device \texttt{ibm\_sydney} was characterized by a large amount of noise and is now deprecated. We used it intentionally to illustrate the performance of the testing procedures in a realistic, noisy environment without error mitigation or error correction. It is important to note that the performance of modern, existing systems is significantly better, and further improvement is expected.

All experiments described in this section are fully reproducible using our code, which is available in \cite{qutest}. 

\subsection{Use case study 1: context-aware unit tests for a three-qubit subroutine}

\begin{figure*}[!t]
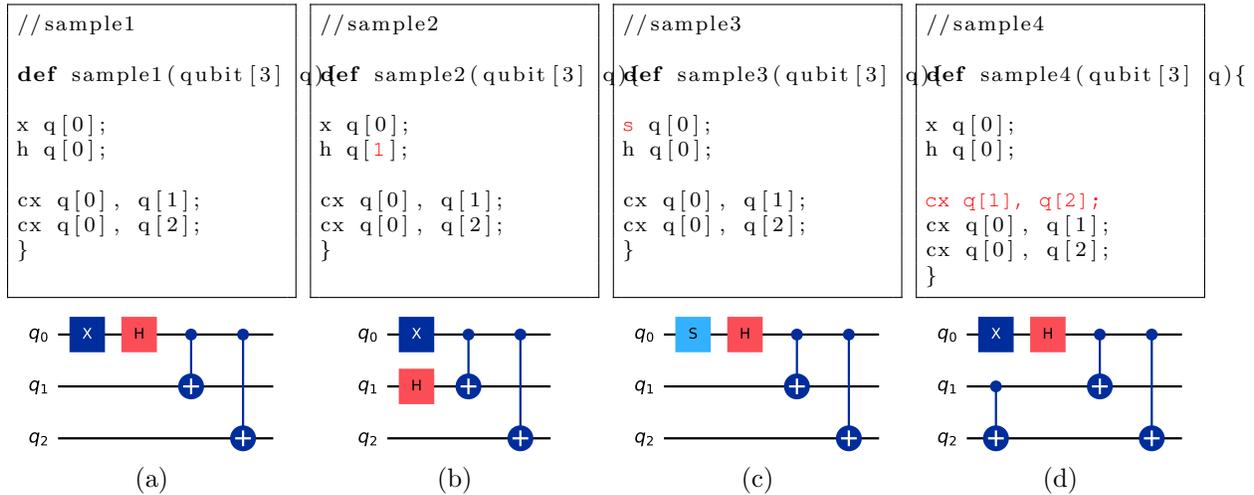

     \centering
\subfloat[]{\usebox{\firstlisting}} \hfill
\subfloat[]{\usebox{\secondlisting}}  \hfill
\subfloat[]{\usebox{\thirdlisting}}  \hfill
\subfloat[]{\usebox{\forthlisting}}\\
\caption{    \label{fig:samples} Samples of source codes written in OpenQASM~3.0 language and corresponding quantum circuit diagrams for three quantum subroutines under testing: (a) the correct three-qubit quantum program, (b) the program with a mutated qubit index for the Hadamard gate, (c) a mutated gate specification (an identity gate instead of a phase shift gate), and d) extra two-qubit gate.}
\end{figure*}

The first set of experiments conducted in this work focuses on unit testing of small three-qubit subroutines. These subroutines (written in OpenQASM~3.0 language) are shown in Figures \ref{fig:samples}a-d  along with the corresponding circuit diagrams. The subroutine labeled \texttt{sample1} in Figure \ref{fig:samples}a is considered as an example of a correct code. This code, when the input state is $\ket{000}$, outputs the Greenberger–Horne–Zeilinger state \cite{10.1119/1.16243}:

\begin{equation}
    \ket{\psi} = \frac{\ket{000}-\ket{111}}{\sqrt{2}}
    \label{eq:bell}
\end{equation}

We also consider three possible mutations of this code, \texttt{sample2}, \texttt{sample3}, and \texttt{sample4}, by introducing three distinct errors (see Figure \ref{fig:samples}b-d): mutation of the qubit label, alteration of the gate specification, and the insertion of an additional two-qubit gate.

With these code samples, we consider two cases of the contextual information: the absence of any contextual information, $\left(C_X=X, C_Y=\mathcal{Y} \right)$, and the scenario where the input state of the subroutine is always a particular state, $\left(C_X=\{\rho_{in}=\ket{000}\bra{000}\}, C_Y=\mathcal{Y} \right)$. The ground truth for the testing experiments using these samples is shown in Table~\ref{tab:gt}. Note, that the outcome for \texttt{sample4} depends on contextual information. Indeed, we can easily verify that when the quantum circuit is initialized by the state $\ket{000}$, the output exactly matches the one given by equation~(\ref{eq:bell}) in both cases. This is because the extra CNOT gate does not introduce any changes when the control qubit is in the state $q_1=\ket{0}$. For all cases when $q_1=\ket{1}$, the outputs of \texttt{sample1} and \texttt{sample4} are different. Therefore, the additional CNOT gate does not alter the specified functionality of the subroutine within the given context. In general, there is no one-to-one correspondence between source code specification and software functionality; different sequences of instructions can achieve the same functionality.

Without specific contextual information, to ensure comprehensive coverage, tests must be conducted in a way that accounts for the entire space of input density matrices $X$ and  space of output density matrices $Y$. This can be achieved using a testing protocol based on the quantum process tomography. The assertions for this protocol are based on the equivalence between the expected and obtained Choi matrices. The context, on the other hand, imposes restrictions on possible input states ans allows for the use of a testing protocol based on quantum state tomography, with assertions that utilize density matrices as arguments.

\begin{figure*}[!t]
    \centering
    \includegraphics[width=0.9\linewidth]{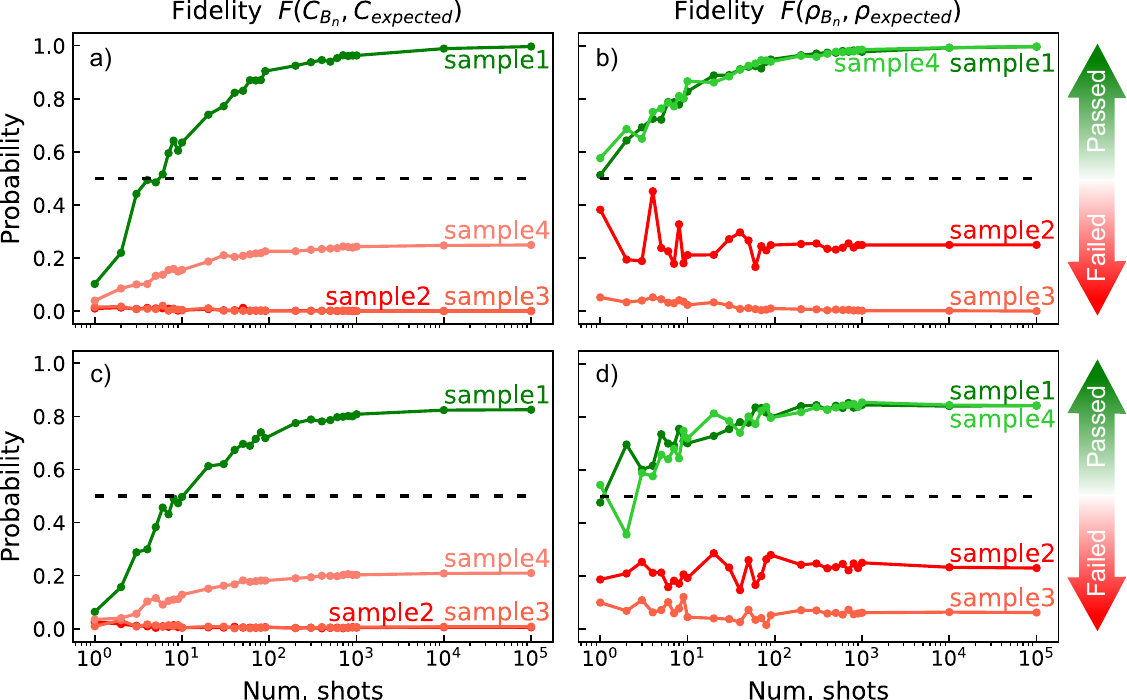}
    \caption{\label{fig:pt_res}Dependence of the probabilistic assertion outcome on number of measurements without contextual information based on the quantum process tomography (panels a and c) and with contextual information based on the quantum state tomography (panels b and d). The experiments have been conducted for the noise-less environment (panels a and b) and in the presence of noise (panels c and d). Note that the ground truth results for the subroutines are also dependent on the contextual information.}
\end{figure*}

The data obtained, shown in Fig. \ref{fig:pt_res}a, indicate that with an extremely small number of shots, false negative outputs are observed for \texttt{sample1}. After approximately ten shots, the probability of \texttt{sample1} passing the test exceeds 0.5 and continues to increase, reaching saturation at a value of 1 after about $10^4$ shots. Conversely, the probability of \texttt{sample2}, \texttt{sample3}, and \texttt{sample4} passing the test remains low, regardless of the number of shots, in contrast to \texttt{sample1}. Thus, these empirical data indicate that, for a small number of measurements, the proposed unit testing methodology based on quantum process tomography is more likely to produce false negative results than false positives. 

\begin{table}[t!]
    \centering
        \caption{Ground truth for the testing experiments}
\begin{tabularx}{0.7\linewidth}{
  | >{\centering\arraybackslash}p{\dimexpr.2\linewidth-2\tabcolsep-1.3333\arrayrulewidth} 
  | >{\centering\arraybackslash}p{\dimexpr.25\linewidth-2\tabcolsep-1.3333\arrayrulewidth}
  | >{\centering\arraybackslash}p{\dimexpr.25\linewidth-2\tabcolsep-1.3333\arrayrulewidth} |  }
  \hline
                     & Without context  &  With context  \\
  \hline
   \texttt{sample1}  & {\leavevmode\color{OliveGreen} [PASSED]} &  {\leavevmode\color{OliveGreen} [PASSED]}  \\
   \texttt{sample2}  & {\leavevmode\color{Red} [FAILED]} &  {\leavevmode\color{Red} [FAILED]}  \\
   \texttt{sample3}  & {\leavevmode\color{Red} [FAILED]} &  {\leavevmode\color{Red} [FAILED]}  \\
   \texttt{sample4}  & {\leavevmode\color{Red} [FAILED]} &  {\leavevmode\color{OliveGreen} [PASSED]} \\
    \hline
\end{tabularx}
    \label{tab:gt}
\end{table}

By including contextual information, we can switch from a testing protocol based on quantum process tomography to a less computationally demanding protocol based on quantum state tomography, without compromising state space coverage. The results of this approach are presented in Figure \ref{fig:pt_res} b for noise-free simulations. Similar to the previous case, this method recovers the expected results with a large number of shots.The dependence on the number of shots is less monotonic than in the previous case, with large fluctuations for small numbers of shots. However, this protocol shows no strong bias toward false negative results for a small number of shots.

We evaluated the performance of each testing protocol by measuring its average wall time per one run. When the protocol based on quantum process tomography is used, the average wall time is $2.13 \times 10^{-3}$  seconds per shot. In contrast, for quantum state tomography with the presence of context information, the average wall time is $4.08 \times 10^{-5}$ seconds per shot. This dramatic difference indicates the importance of the contextual information in the quantum unit testing. This difference becomes particularly significant when aiming to achieve reliable test outcomes for large-scale systems.

The same results, but with noise included in the simulations, are shown in Fig. \ref{fig:pt_res} c and d. In both cases, at a large number of shots the test still produces correct outputs for the cases considered, but with reduced confidence for \texttt{sample1}, which reaches a maximal value of 0.8. 

\subsection{Use case study 2: unit tests for Shor's algorithm}

The second case study aims to illustrate the performance of the proposed testing methodologies when applied to a real-world application. In particular, we consider quantum phase estimation (QPE), which is an important part of Shor's algorithm \cite{shor1999polynomial, buchmann2024introduction}. Shor's algorithm factors an integer $N$ by reducing the problem to finding the order of an integer $a$ modulo $N$, where the order is the smallest positive integer $r$ such that $a^r \equiv 1 \mbox{ mod } N$.  The quantum circuit diagram representing QPE is shown in Figure~\ref{fig:shor} and itself consists of two distinct subroutines: one performs a controlled unitary operation (Subroutine 1), and the other implements the inverse quantum Fourier transform (Subroutine 2). Subroutine 1 has a classical parameter 
$\theta_j$ and is reused multiple times with different values of the parameter.

Both subroutines have been implemented using the \texttt{Qiskit} framework \cite{qiskit}. Their detailed description can be found in the literature (see for instance Refs. \cite{shor1999polynomial, buchmann2024introduction}) and in the Supplementary Materials (Section 1 and 2) accompanying this work.

For Subroutine 1, as specified by the algorithm's functional requirements, the input state is always the same, $\rho_{in}$, which defines the context: $\left( C_X=\{\rho_{in} \}, C_Y=\mathcal{Y} \right)$. This information allows us to apply the testing protocol based on quantum state tomography to ensure full state space coverage. The correct implementation of Subroutine 1 also implies that the state $\rho_{in}$ is supposed to be an eigenvector of the unitary operation realized by this subroutine for any values of the parameter
$\theta$. The unitary matrix acting on its eigenvector must return the same eigenvector multiplied by a global phase factor. Therefore, the state vector should remain unchanged after the action of Subroutine 1 if the implementation is error-free. We use this fact as the basis for designing a unit test, applying a testing protocol based on state tomography and using an assertion based on the equality of the density matrices for the input and output states, $\rho_{in} = \rho_{out}$. The assertion is evaluated by computing the fidelity between the input and output states, according to equation~(\ref{eq:fid}).

\begin{figure*}[!t]
\centering
\includegraphics[width=0.9\linewidth]{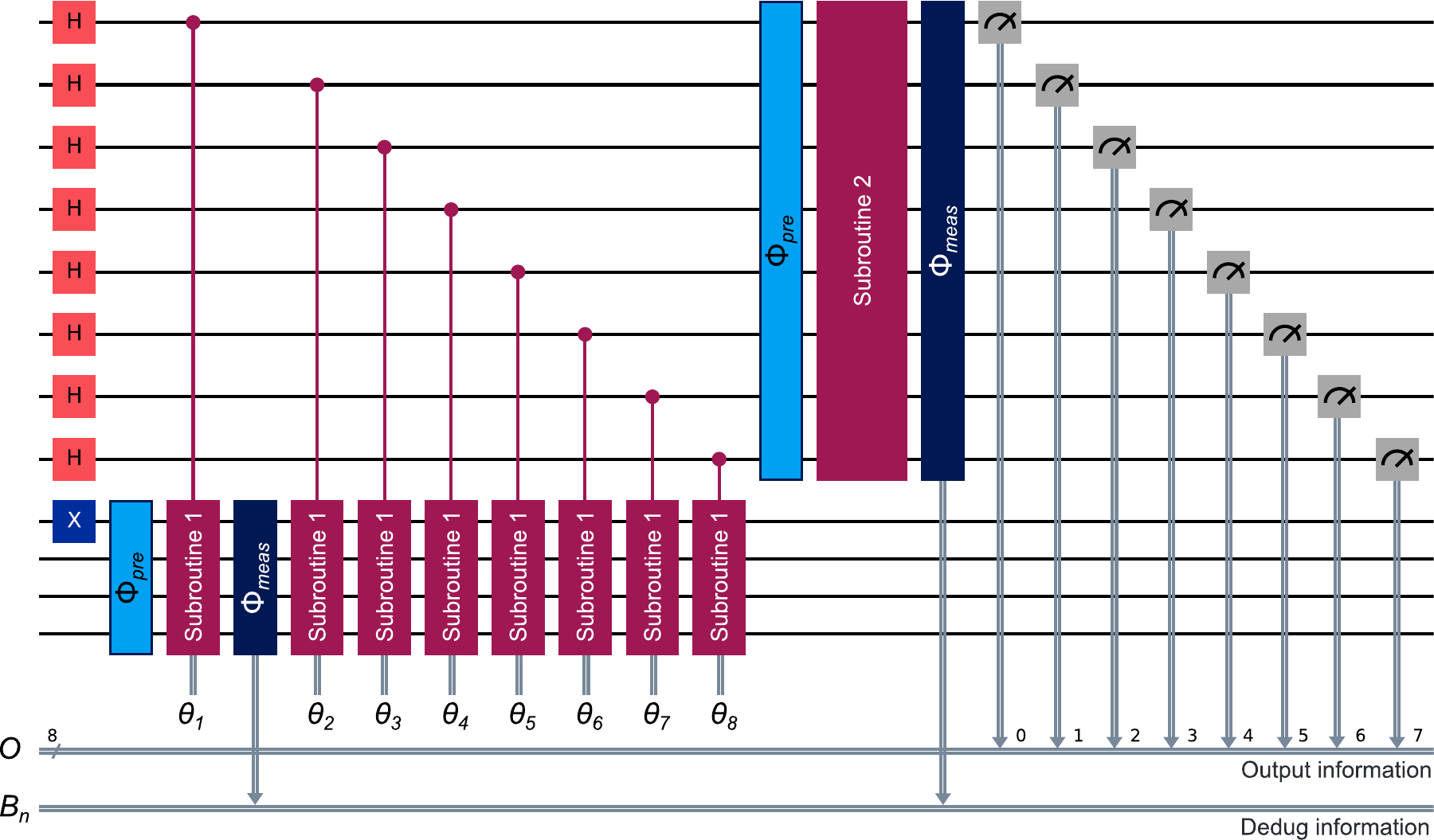}
    \caption{Circuit diagram of a quantum program that implements the phase estimation algorithm, which is a key component of Shor's algorithm (here Subroutine 1 implements a parameterized unitary rotation, while Subroutine 2 performs the inverse quantum Fourier transform). The quantum channels $\Phi_{pre}$ and $\Phi_{meas}$ are only applied during testing.}
    \label{fig:shor}
\end{figure*}

\begin{figure}[t]
    \centering
    \includegraphics[width=0.5\linewidth]{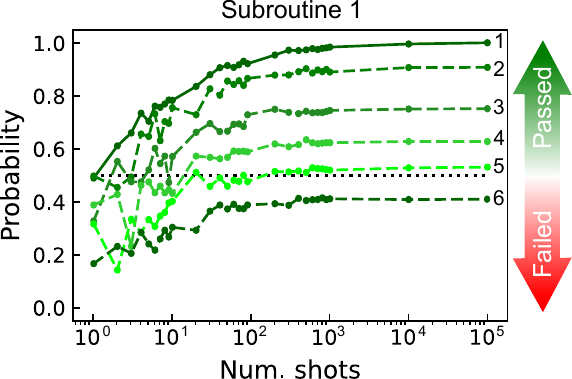}
    \caption{\label{fig:shor_results1}Dependence of the probabilistic assertion outcome on number of shots for a correct implementation of Subroutine 1. The assertion has been evaluated using the testing protocol based on the quantum state tomography. The results are obtained for a noise-free environment (solid line) and with noise models (dashed lines). The numbers at the curves indicate the noise models used in the simulations: 1 represents the noiseless case; 2–5 correspond to bit-flip noise applied to each qubit and operation, with probabilities of 0.001, 0.003, 0.005, and 0.007, respectively; 6 denotes the noise model from \texttt{ibm\_sydney}.}
    \label{fig:shor}
\end{figure}

The results of this assertion evaluation are shown in Fig. \ref{fig:shor_results1}. For the noise-free environment, the dependence of the output on the number of shots is similar to the previous case when quantum state tomography was applied to a three-qubit subroutine: for a small number of shots, the probability is close to 0.5, then gradually increases and eventually saturates close to 1.0, confirming that the subroutine has passed the test. However, when accounting for the realistic noise levels characteristic of \texttt{ibm\_sydney}, the outcomes become unreliable even with a large number of shots (curve 6 in Fig. \ref{fig:shor_results1}). This pronounced difference from the three-qubit case - despite using the same simulation environment - can be attributed to the increased number of qubits. QPE requires many gates and many qubits for any realistic application, and thus requires very low levels of noise. It is not intended as a NISQ algorithm and not suspected to be executable on currently-available noisy hardware. Thus we expect QPE \textit{should} fail our tests when the testing protocol based on the quantum state tomography is simulated with realistic noise models. 

For Subroutine 1 we also considered simpler noise models based on the Pauli quantum error channel, where the qubit flip operation $X$ is applied to each qubit with varying probabilities (in Fig. \ref{fig:shor_results1} probabilities 0.001, 0.003, 0.005, 0.007 correspond to curves 2, 3, 4 and 5 respectively). These additional models with varying error probabilities demonstrate that, at a large number of shots, the probability of the assertion depends almost linearly on the qubit flip probability, spanning outcomes from the noiseless case to those generated by a realistic noise model for these particular values. 

Thus, in testing protocols based on quantum tomography, a high level of hardware errors introduces uncertainty into test outcomes and, at extreme levels, can result in false negatives. This underscores the impact of hardware noise on testing reliability and highlights the critical importance of fault-tolerant quantum computers and the implementation of error correction mechanisms in quantum unit testing.

Subroutine 2 implements the inverse quantum Fourier transform algorithm for 8 qubits (see Supplementary Materials, Section 2, for a detailed description). Note that previously, full quantum process tomography was applied to test the quantum Fourier transform for three qubits \cite{10.1063/1.1785151}. Due to the exponential complexity in terms of qubit count, it is difficult to apply this protocol to systems beyond that scale. Luckily, for the QPE algorithm, the use of the quantum Fourier transform is also supplemented with nontrivial context information - namely, the output state undergoes destructive measurements immediately after the action of this subroutine: $\left( C_X=X, C_Y=\{\mbox{tr} \left(O \rho \right)\} \right)$, where $O$ is the operator corresponding to the projective measurements for all qubits in the computational basis set. This context implies that we do not need to obtain information on all elements of the density matrix corresponding to the output state, but only those relevant to the amplitudes of the projective measurement outcomes. Therefore, it is sufficient to use testing protocols based on statistical tests, avoiding expensive quantum tomography protocols. 

\begin{figure}[t]
    \centering
    \includegraphics[width=0.5\linewidth]{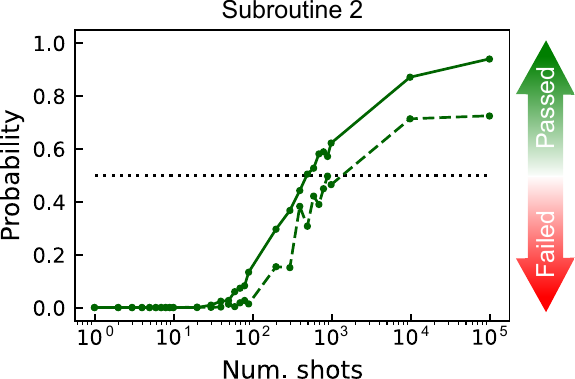}
    \caption{\label{fig:shor_results2}Dependence of the probabilistic assertion outcome on number of shots for a correct implementation of Subroutine 2. The assertion has been evaluated using the testing protocol based on the Pearson's $\chi^2$ test and p-value. The results are obtained for a noise-free environment (solid line) and with noise model from \texttt{ibm\_sydney} (dashed line).}
\end{figure}

In this example, we employ a testing protocol based on the Pearson's $\chi^2$ test, using the p-value to assess how well the measured outcome count frequencies align with the expected frequencies. Moreover, for this test, we do not intend to achieve full state space coverage like in all previous cases; instead, we perform the test for only one value of the input state and one particular value of the parameter $\theta$. For the quantum Fourier transform, a reliable test can be based on the Fourier transform of a delta function, the output of which can be computed analytically. One way to introduce a delta function is by initializing all qubits in the input quantum register to zero and then applying a Hadamard gate to the first qubit. This operation constitutes the sole element of the set $\{ \Phi_{pre}^j \}$. For this testing protocol, the set $\{ \Phi_{meas}^j \}$ also contains a single element -- a quantum channel that performs projective measurements in the computational basis. The outcomes of the experiments (count frequencies for each state) are presented in the Supplementary Materials (Section 3). The results of the probabilistic assertion evaluation, shown in Fig. \ref{fig:shor_results2}, indicate that a reliable outcome can only be achieved after at least $10^4$ shots - significantly more than what is required when using the protocol based on quantum tomography. On the other hand, for this particular subroutine, the test protocol based on the $\chi^2$-test produces correct output even in the presence of realistic hardware noise. The presence of noise only reduces the level of certainty of the results - even with a large number of shots, the probability of the test passing does not exceed 0.8. The employed testing protocol clearly requires less computational time, as it involves fewer measurements compared to quantum state tomography.

\section{Discussion}

\subsection{Answers to the research questions}

Here, we briefly summarize the answers to the research questions addressed in this study.

\textbf{RQ1:} Quantum subroutines, as well as quantum programs, can be modeled as parametrized quantum channels, which enables the use of quantum process tomography and quantum state tomography to verify their expected functionality.

\textbf{RQ2:} As demonstrated in this work, unit tests for quantum software can include various assertion propositions with different types of arguments. These types of arguments include Choi matrices for quantum processes, density matrices, and density matrix functionals, such as observables and quantum state properties. The particular choice of arguments, along with the methodology used to retrieve them from the measurement data and to evaluate the assertion proposition, determines a specific testing protocol. In this work, we have identified testing protocols based on quantum process tomography, standard quantum state tomography, classical shadow tomography, Pearson's
 $\chi^2$-test, and single-shot measurements. The choice of testing protocol depends on the specific case and requirements for test precision, the availability of computational resources (including the number of qubits, computational time, and cost), the reliability of the hardware (such as noise levels and the presence of error correction), and other relevant factors. 

\textbf{RQ3:} The computational complexity of unit tests can be reduced by breaking the code into relatively small, logically meaningful subroutines and specifying the context for each one. The context provides additional information on how the quantum subroutine is used in the application. This information influences the choice of testing protocol and can significantly reduce the computational complexity of unit tests by lowering the dimensionality of the input and output state spaces, as well as relaxing the accuracy requirements. Without context, unit tests with the greatest coverage are based on quantum process tomography.

\textbf{RQ4:} The obtained results demonstrate that different testing protocols exhibit varying sensitivities to noise. Additionally, quantum programs with a larger number of qubits are more susceptible to errors induced by noise. In general, hardware noise diminishes the certainty of the test outcomes, even for a large number of shots.

\subsection{Trade-offs}

\begin{figure*}[!t]
    \centering
    \includegraphics[width=0.87\linewidth]{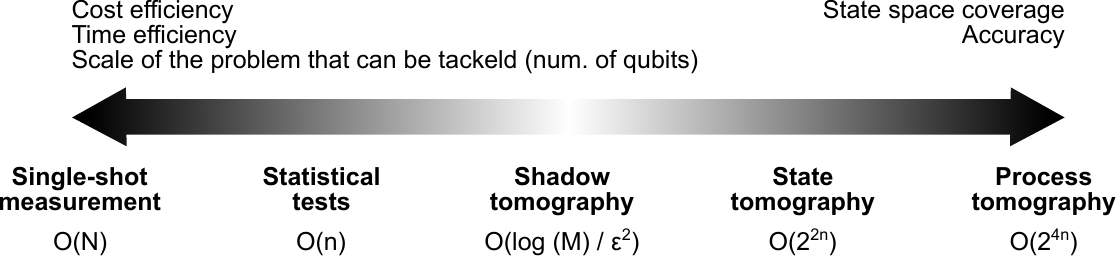}
    \caption{ \label{fig:trade} Trade-offs between state space coverage, accuracy, time, and cost efficiency for different quantum unit testing protocols. The diagrams also show the big-O-notation for asymptotic complexity (in terms of number of measurements) as a function of $n=N \cdot \alpha$, where $N$ is a number of qubits and $\alpha$ is a number of shots.}
\end{figure*}

As demonstrated in this work and illustrated in Figure \ref{fig:trade}, the general approach to black-box quantum unit testing is characterised by a trade-off between accuracy and state space coverage, on the one hand, and cost, time efficiency, and problem scale on the other. Different testing protocols and corresponding choices of assertion arguments offer different combinations of these attributes. When designing quantum unit tests, it is the engineer's responsibility to assess the required testing accuracy and available resources in order to select the most appropriate testing protocol. This work has shown that the choice of testing protocol -- and the associated trade-offs -- is strongly influenced by the available contextual information regarding how the quantum subroutine is used within the project. Such information may reduce the need for additional measurements and tests, and must be carefully considered when developing the testing strategy.
 
\subsection{Related works}

To date, there are few publications focused on testing in quantum software, which have been reviewed in \cite{garcia2023quantum}. To formally verify the correctness of quantum programs, a Floyd–Hoare logic for quantum programs has been developed in  \cite{10.1145/2049706.2049708}. This is an adaptation of classical Floyd–Hoare logic to quantum computing, a formal system with a set of logical rules for rigorously reasoning about the correctness of computer programs written in a simple imperative programming language. The key notion of this formal system is a Hoare triple  $\{P\}C\{Q\}$, which describes the change in the state of the computer memory, where $P$ and $Q$ are the precondition and postcondition assertions, and $C$ is a command or a program. Transitioning from formal reasoning to empirical testing gives rise to one of the most commonly used patterns in classical unit testing, known as ``arrange-act-assert" \cite{okken2022python}. This implies that the general structure of a unit test consists of three components: setting up the initial state of the object to be tested (arrange), executing the test on the object within the previously set-up environment (act), and assessing whether the output matches the expected results (assert). 
A way to evaluate assertions for quantum programs has been proposed in \cite{10.1145/3428218} using non-destructive projection measurements. The advantage of this approach is that, in theory, unit testing can be performed during runtime, when the quantum program continues to function uninterruptedly. However, this approach cannot be applied as a universal methodology for isolated black-box unit tests, as it provides limited information and does not cover all possible input-output cases for a given quantum subroutine. In this work, we examine an alternative approach based on destructive measurements and quantum tomography, where the non-destructive projections can be seen as a special case.

\section{Conclusion}

In this work, we propose a unified theoretical framework for unit testing quantum subroutines that integrates a range of testing protocols, such as quantum process tomography, quantum state tomography, Pearson's $\chi^2$ test, and others. The choice of a particular protocol depends on the requirements for accuracy, available computational resources (classical and quantum), and accessible contextual information. For a quantum subroutine, the contextual information can constrain both the dimensionality of the input state space and the possible uses of the output quantum state, as only certain projections of the output state may be further utilised in a larger project.

It has been established that quantum unit tests are inherently probabilistic for two reasons: first, the tomographic reconstruction of the density matrix is a probabilistic process governed by a likelihood function; second, the distinguishability of quantum states or distribution functions, given a limited number of system copies, is also characterised by an inherent probability of error. The results of a series of numerical experiments and simulations carried out in this work indicate that reliable test outcomes for probabilistic assertions require a large number of runs using identical copies of the systems. Hardware noise decreases the reliability and certainty of test outcomes.

The topic of unit testing in classical computation is rich and deep, and extending this process to quantum computation opens up several promising directions for future research, and raises many questions which will need serious consideration as quantum computing makes the transition from pure research to practical software development. In particular, future work will focus on developing a unit testing framework that leverages probabilistic contextual information through Bayesian inference, as well as on incorporating non-trivial prior knowledge of the density matrix. Additionally, we aim to evaluate the accuracy of tomography-based tests that uses randomized sampling techniques, instead of relying solely on tomographically complete measurement sets

\section{References}

\bibliographystyle{iopart-num}
\bibliography{main}

\setcounter{section}{0}
\setcounter{figure}{0}
\setcounter{table}{0}

\lstdefinestyle{mystyle}{
  backgroundcolor=\color{backcolour},   commentstyle=\color{codegreen},
  keywordstyle=\color{magenta},
  numberstyle=\tiny\color{codegray},
  stringstyle=\color{codepurple},
  basicstyle=\ttfamily\scriptsize,
  breakatwhitespace=false,         
  breaklines=true,                 
  captionpos=b,                    
  keepspaces=true,                 
  numbers=left,                    
  numbersep=5pt,                  
  showspaces=false,                
  showstringspaces=false,
  showtabs=false,                  
  tabsize=2,
  showlines=true,
}

\lstset{style=mystyle}

\title{Supplementary Materials: Context-Aware Unit Testing for Quantum Subroutines}

\author{Mykhailo Klymenko$^1$, Thong Hoang$^2$, Sam Wilkinson$^1$, Bahar Goldozian$^1$, Suyu Ma$^1$, Xiwei Xu$^{2,3}$, Qinghua Lu$^{2,3}$, Muhammad Usman$^{1,4}$, Liming Zhu$^{2,3}$}

\address{$^1$Data61, CSIRO, Research Way, Clayton 3168, VIC, Australia}
\address{$^2$Data61, CSIRO, Level 5/13 Garden St, Eveleigh 2015, NSW, Australia}
\address{$^3$University of New South Wales, Sydney 2052, NSW, Australia}
\address{$^4$School of Physics, The University of Melbourne, Parkville 3010, VIC, Australia}

\maketitle

All examples and code listings referenced in this work can be found in \url{https://github.com/klymenko-code/qutest}. In this work, all tests of Shor’s algorithm have been implemented using the factorization of the number 9 as a simple example. Shor’s algorithm integrates both classical and quantum components. The quantum part is realized through the quantum phase estimation algorithm and consists of two subroutines, which are described below.

\section{Subroutine 1}

The source code of Subroutine 1, written in \texttt{Qiskit} - a high-level quantum programming language - is shown in Fig. \ref{fig:u_a}.

\begin{figure}[!h]
    \centering
    \begin{lstlisting}[language=Python]
from math import gcd, floor, log
import numpy as np
from qiskit.circuit.library import UnitaryGate

    
def mod_mult_cirquit(circ, theta, N, a=2):
    """Subroutine 1

    :param circ:  input quantum register
    :param theta: input classical argument, orders of the unitary matrix, theta=1..8
    :param N:     input parameter, the number to factorize
    :param a:     a random integer 2 <= a < N such that gcd(a,N) > 1
    :return:      qiskit circuit for Subroutine 1"""

    qubit = circ.qubits[theta]
    target = circ.qubits[-1]._register
    circ.h(theta)
    b = pow(a, 2 ** theta, N)

    if gcd(b, N) > 1:
        print(f"Error: gcd({b},{N}) > 1")
    else:
        n = floor(log(N - 1, 2)) + 1
        U = np.full((2 ** n, 2 ** n), 0)
        for x in range(N): U[b * x % N][x] = 1
        for x in range(N, 2 ** n): U[x][x] = 1
        G = UnitaryGate(U)
        G.name = f"M_{b}"

    circ.compose(G.control(), qubits=[qubit] + list(target), inplace=True)

    return circ
\end{lstlisting}
    \caption{Source code for Subroutine 1}
    \label{fig:u_a}
\end{figure}

This representation was chosen for its compactness achieved by high-level abstractions offered by the programming language. While this code can be translated into QASM instructions or circuit diagrams, such representations would require many lines of text due to the large number of primitive instructions/gates. Ultimately, the specific set of instructions for this subroutine is not unique and depends on the instruction set architecture of the backend and the level of optimization.

Subroutine 1 implements as a controlled unitary rotation $U_a^{\theta}$ acting on four qubits from the register $T$ (see the circuit diagram in Fig. 6 of the main text). For each value of 
$\theta$, this unitary matrix corresponds to a permutation matrix for quantum states.  In the phase estimation algorithm, each usage of this subroutine $\theta$ is

In this work, Subroutine 1 has been tested with the control qubit initialized in the state $\vert 1 \rangle$, and for a specific parameter value $\theta=1$ and $a=2$. Under these conditions, the unitary operation is represented by the following matrix:

\begin{equation}
   U_{a=2}^{\theta=1} =  \left(  \begin{array}{cccccccccccccccc}
      1 & 0 & 0 & 0 & 0 & 0 & 0 & 0 & 0 & 0 & 0 & 0 & 0 & 0 & 0 & 0 \\  [-0.27cm]
      0 & 0 & 0 & 0 & 0 & 1 & 0 & 0 & 0 & 0 & 0 & 0 & 0 & 0 & 0 & 0 \\ [-0.27cm]
      0 & 1 & 0 & 0 & 0 & 0 & 0 & 0 & 0 & 0 & 0 & 0 & 0 & 0 & 0 & 0 \\  [-0.27cm]
      0 & 0 & 0 & 0 & 0 & 0 & 1 & 0 & 0 & 0 & 0 & 0 & 0 & 0 & 0 & 0 \\  [-0.27cm]
      0 & 0 & 1 & 0 & 0 & 0 & 0 & 0 & 0 & 0 & 0 & 0 & 0 & 0 & 0 & 0 \\  [-0.27cm]
      0 & 0 & 0 & 0 & 0 & 0 & 0 & 1 & 0 & 0 & 0 & 0 & 0 & 0 & 0 & 0 \\  [-0.27cm]
      0 & 0 & 0 & 1 & 0 & 0 & 0 & 0 & 0 & 0 & 0 & 0 & 0 & 0 & 0 & 0 \\  [-0.27cm]
      0 & 0 & 0 & 0 & 0 & 0 & 0 & 0 & 1 & 0 & 0 & 0 & 0 & 0 & 0 & 0 \\  [-0.27cm]
      0 & 0 & 0 & 0 & 1 & 0 & 0 & 0 & 0 & 0 & 0 & 0 & 0 & 0 & 0 & 0 \\  [-0.27cm]
      0 & 0 & 0 & 0 & 0 & 0 & 0 & 0 & 0 & 1 & 0 & 0 & 0 & 0 & 0 & 0 \\  [-0.27cm]
      0 & 0 & 0 & 0 & 0 & 0 & 0 & 0 & 0 & 0 & 1 & 0 & 0 & 0 & 0 & 0 \\  [-0.27cm]
      0 & 0 & 0 & 0 & 0 & 0 & 0 & 0 & 0 & 0 & 0 & 1 & 0 & 0 & 0 & 0 \\  [-0.27cm]
      0 & 0 & 0 & 0 & 0 & 0 & 0 & 0 & 0 & 0 & 0 & 0 & 1 & 0 & 0 & 0 \\  [-0.27cm]
      0 & 0 & 0 & 0 & 0 & 0 & 0 & 0 & 0 & 0 & 0 & 0 & 0 & 1 & 0 & 0 \\  [-0.27cm]
      0 & 0 & 0 & 0 & 0 & 0 & 0 & 0 & 0 & 0 & 0 & 0 & 0 & 0 & 1 & 0 \\  [-0.27cm]
      0 & 0 & 0 & 0 & 0 & 0 & 0 & 0 & 0 & 0 & 0 & 0 & 0 & 0 & 0 & 1 \\ 
   \end{array} \right)
\end{equation}

The order of elements in the matrix corresponds to the elements of the computational basis set of four qubits sorted lexicographically.

\section{Subroutine 2}

The subroutine for the inverse quantum Fourier transform has been taken from the \texttt{Qiskit} circuit library.  The code listing for this subroutine used in this work is shown in Fig. \ref{fig:iqft}.

\begin{figure}[!h]
    \centering
    \begin{lstlisting}[language=Python]
from qiskit.circuit.library import QFT

    
def fourier(circ):
    """
    Subroutine 2
    
    :param circ: input quantum register
    :return:        qiskit circuit for Subroutine 1
    """
    
    control = circ.qubits[0]._register

    circ.compose(
        QFT(len(control), inverse=True),
        qubits=control,
        inplace=True)

    return circ
\end{lstlisting}
    \caption{Source code for Subroutine 2}
    \label{fig:iqft}
\end{figure}

\begin{figure}[h!]
    \centering
    \includegraphics[width=0.8\linewidth]{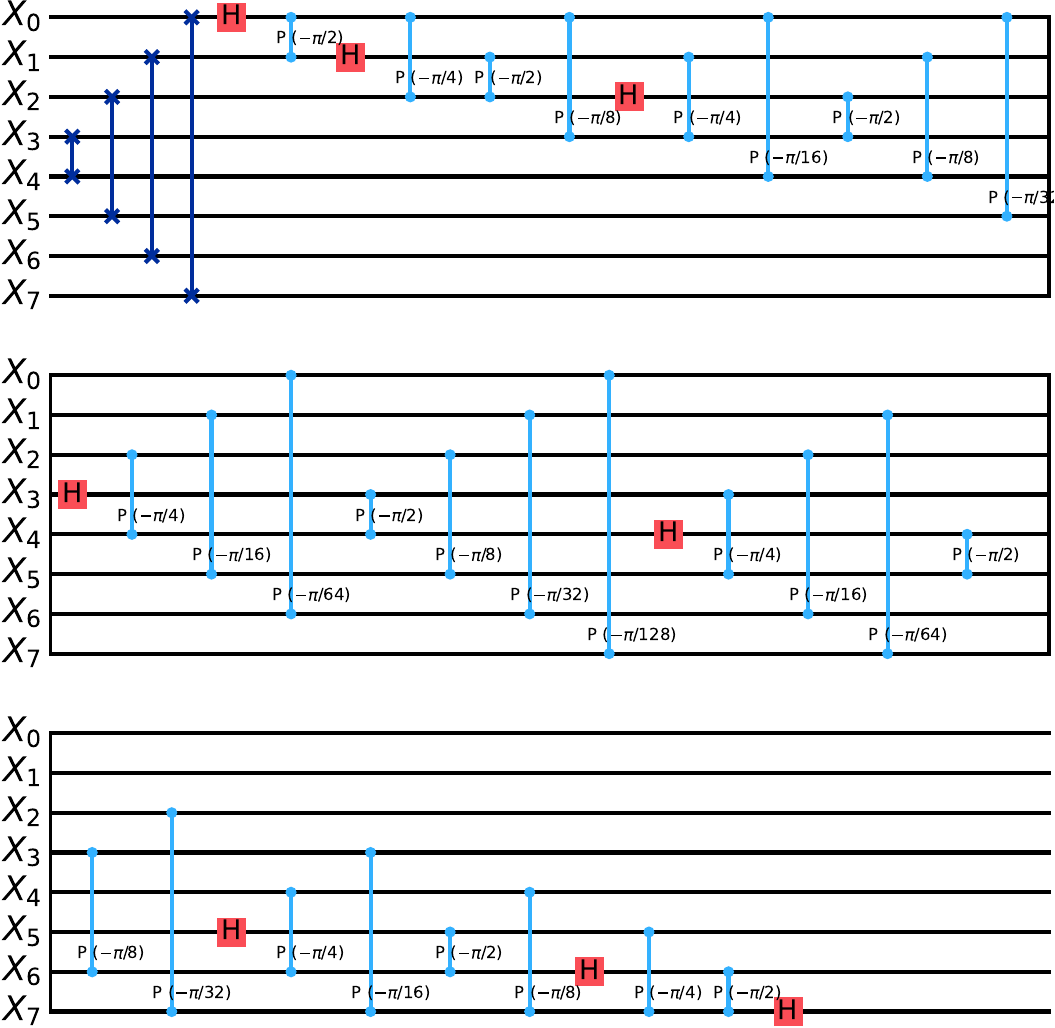}
    \caption{ \label{fig:iqft1} Circuit diagram of the library function QTF implementing the inverse quantum Fourier transform }
\end{figure}

For the reference, we also show the circuit diagram for the QFT function from the \texttt{Qiskit} library in Fig. \ref{fig:iqft1}, and a detailed description of its design, functionality, and this particular implementation can be found in the literature.
As illustrated in Fig. \ref{fig:iqft1}, even for an 8-qubit system, the circuit consists of numerous single- and two-qubit gates. If programmed manually, this complexity makes the subroutine prone to human-induced errors, highlighting the need for unit testing.

\section{Results of experiments for Subroutine 2}

\begin{figure}[h!]
    \centering
    \includegraphics[width=0.8\linewidth]{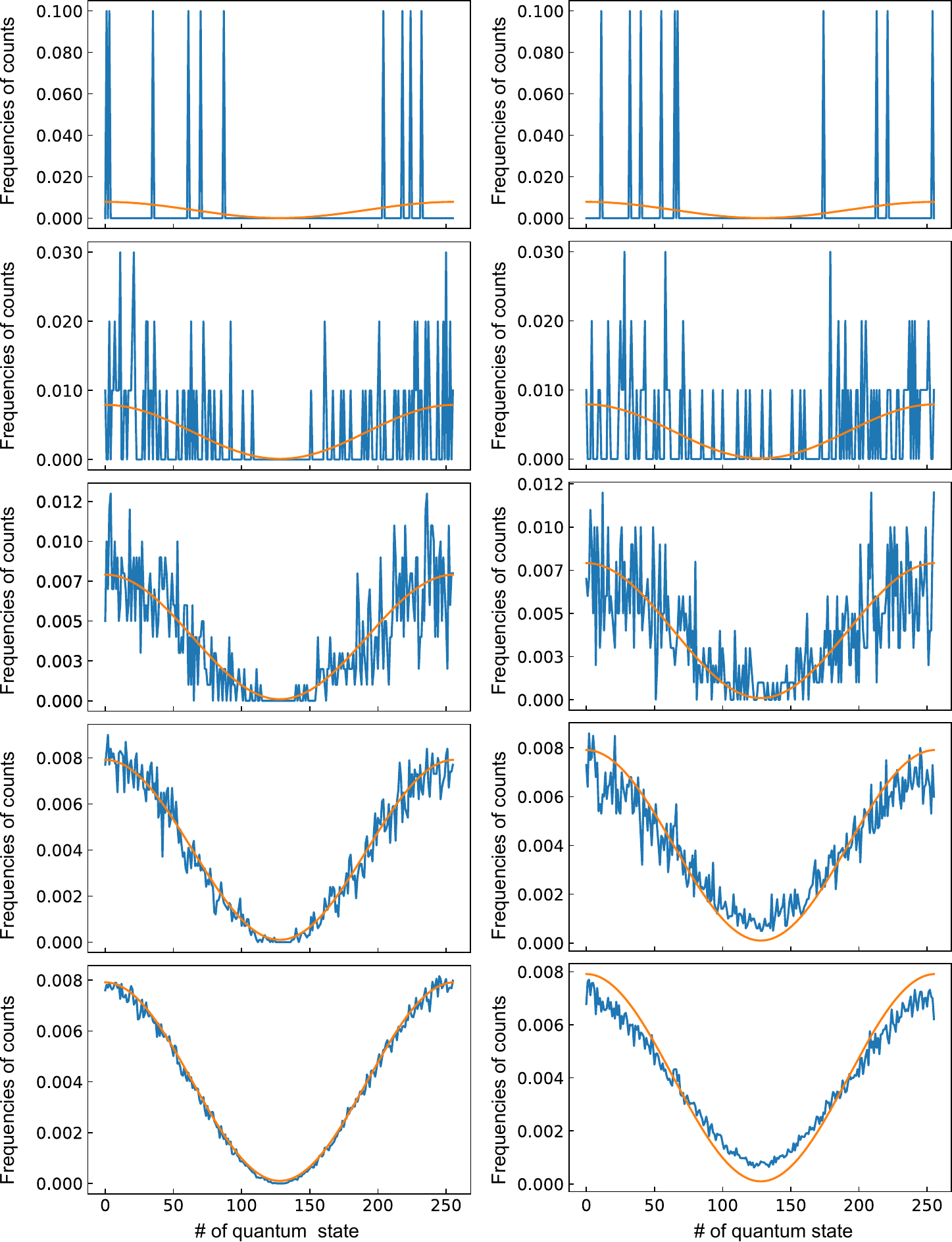}
    \caption{ \label{fig:iqft_red} Expected (orange line) vs. experimentally obtained (blue line) counts per quantum state, shown without noise (left panels) and with noise (right panels). Each row of panels corresponds to an increasing number of shots: $10^1$, $10^2$, $10^3$, $10^4$, $10^5$, from top to bottom.}
\end{figure}

In Fig. \ref{fig:iqft_red}, we present intermediate raw data obtained during the unit testing of Subroutine 2. These data represent measurement outcomes of qubit states in the computational basis, which are used to evaluate the 
$R^2$ metric discussed in the main text. The mean value of the expected number of counts follows a cosine function, which corresponds to the Fourier transform of a delta function.

\end{document}